\documentclass[english,aps,preprint,showpacs]{revtex4}
\usepackage[T1]{fontenc}
\usepackage[latin9]{inputenc}
\usepackage{amssymb}
\usepackage{graphicx}

\makeatletter

\usepackage{babel}
\makeatother
\begin{document}

\preprint{This line only printed with preprint option}

\title{Fermi breakup and the Statistical Multifragmentation Model}

\author{B.V.\ Carlson$^{1}$}

\author{R.\ Donangelo$^{2,3}$}

\author{S.R.\ Souza$^{2,4}$}

\author{W.G.\ Lynch$^{5}$}

\author{A.W.\ Steiner$^{5}$}

\author{M.B.\ Tsang$^{5}$}

\affiliation{$^{1}$Departamento de F\'{\i}sica, Instituto Tecnológico de Aeronáutica
- CTA, 12228-900\\
 São José dos Campos, Brazil}

\affiliation{$^{2}$Instituto de F\'{\i}sica, Universidade Federal do Rio de
Janeiro Cidade Universitária, \\
 CP 68528, 21941-972, Rio de Janeiro, Brazil}

\affiliation{$^{3}$ Instituto de F\'{\i}sica, Facultad de Ingenier\'{\i}a Universidad
de la República, Julio Herrera y Reissig 565, 11.300 Montevideo, Uruguay }

\affiliation{$^{4}$Instituto de F\'{\i}sica, Universidade Federal do Rio Grande
do Sul\\
 Av. Bento Gonçalves 9500, CP 15051, 91501-970, Porto Alegre, Brazil}

\affiliation{$^{5}$ Joint Institute for Nuclear Astrophysics, National Superconducting
Cyclotron Laboratory, and the Department of Physics and Astronomy,
Michigan State University, East Lansing, MI 48824, USA}

\begin{abstract}
We demonstrate the close similarity of a generalized Fermi breakup
model, in which densities of excited states are taken into account,
to the microcanonical statistical multifragmentation model used to
describe the desintegration of highly excited fragments of nuclear
reactions. 
\end{abstract}


\maketitle

\section{Introduction}

Both the Fermi breakup (FBM) and the statistical multifragmentation
(SMM) models provide prescriptions for calculating mass and charge
distributions and multiplicities of the fragments emitted in the breakup
of an excited nuclear system. Yet they would seem to be very different
models. They are formulated in diferent terms and usually applied
in very different regions of mass and excitation energy. We will show
that they are much more closely related than they might appear to
be at first glance.

The FBM was originally proposed as a means of calculating the multiplicities\cite{Fermi1}
and angular distributions\cite{Fermi2} of pions and antiprotons produced
in high-energy collisions of cosmic-ray protons with nucleons in the
atmosphere. It was found to be quite successful in this respect\cite{Lord}.
It was later applied to the statistical decay of light fragments of
proton-induced spallation reactions\cite{Lepore,Rozental,Epherre1}
and is now included as the preferred option for the equilibrium statistical
decay of light fragments in widely-used nuclear reaction/transport
codes, such as FLUKA\cite{Ferrari} and GEANT4\cite{geant}. A variant
called the phase space model, which partially takes incomplete equilibration
into account, plays an important role in the analysis of experimental
multi-particle fragmentation spectra in light-ion reactions\cite{Delbar,Rubens1,Bohlen,Rubens2,Rubens3}.
In the context of nuclear reactions, the FBM usually assumes that
fragments are emitted in their ground states or in (almost) particle-stable
excited states and is formulated directly in terms of a phase-space
integral limited only by the constraints of linear momentum and energy
conservation.

The SMM is used to describe the decay of highly-excited fragments
of heavy-ion or spallation reactions. It assumes thermal equilibrium
and thus allows for the emission of particle-unstable excited fragments
consistent with that equilibrium. It has been widely compared to experimental
data and found to reproduce them reasonably well\cite{Bill1,Bill2,Bill3,Bill4}.
Although many versions of the SMM have been proposed over the years\cite{Raul1,Raul2,Raul3,Sneppen,Botvina1,Botvina2,DasGupta1,Sergio1,Sergio2},
it was first developed systematically in Refs. \onlinecite{Raul1,Raul2,Raul3}.
The SMM is normally formulated in terms of a statistical partition
function, be it microcanonical, canonical or grand canonical. The
most appropriate of these is the microcanonical partition function,
for which charge, mass number and energy are strictly conserved. The
canonical and grand canonical partition functions are useful for deriving
analytical or semi-analytical expressions that would be impossible
to obtain in the microcanonical formulation or, when fluctuations
are small, to simplify calculations.

In the following we will demonstrate how a model very similar to the
SMM can be obtained from an appropriately generalized FBM. We will
then discuss the differences between the SMM and the generalized
FBM derived here and consider possible directions of future work.

\section{A Generalized FBM}

A justification of the FBM would begin with the transition rate $\gamma_{0\rightarrow n}$
from a state $0$ to the states of a configuration $n$, \begin{equation}
\gamma_{0\rightarrow n}=\frac{2\pi}{\hbar}\left|\tau_{0\rightarrow n}\right|^{2}\omega_{n}\,,\end{equation}
 where $\omega_{n}$ is the density of (linear momentum and energy
conserving) states of the configuration per unit energy and, most
importantly, the transition matrix element $\tau_{0\rightarrow n}$
is assumed to be independent of the individual momenta of the final
configuration. In the steady state, the probability of producing a
given configuration $n$ can then be calculated as \begin{equation}
P_{n}=\frac{\gamma_{0\rightarrow n}}{\sum_{m}\gamma_{0\rightarrow m}}\,,\end{equation}
 where the sum in the denominator runs over all possible configurations.
When the transition matrix element is configuration independent as
well, \begin{equation}
\tau_{0\rightarrow n}=\tau_{0}\,,\end{equation}
 then the probability of a configuration depends only on its density
of final states,\begin{equation}
P_{n}=\frac{\frac{2\pi}{\hbar}\left|\tau_{0}\right|^{2}\omega_{n}}{\sum_{m}\frac{2\pi}{\hbar}\left|\tau_{0}\right|^{2}\omega_{m}}=\frac{\omega_{n}}{\sum_{m}\omega_{m}}\,.\end{equation}
 This is assumed to be the case in the FBM.

In applications of the FBM to nuclear decay\cite{Ferrari,geant},
the phase-space integral that determines the density of final states
of a configuration of $n$ fragments is usually written as\begin{eqnarray}
\omega_{n} & = & \prod_{l=1}^{k}\frac{1}{N_{l}!}\,\left(\frac{V_{n}}{\left(2\pi\hbar\right)^{3}}\right)^{n-1}\prod_{j=1}^{n}g_{j}\int\prod_{j=1}^{n}d^{3}p_{j}\,\delta\left(\sum_{j=1}^{n}\vec{p}_{j}\right)\,\label{FBM}\\
 &  & \qquad\qquad\qquad\qquad\times\delta\left(\varepsilon_{0}-B_{0}-E_{c0}-\sum_{j=1}^{n}\left(\frac{p_{j}^{2}}{2m_{j}}-B_{j}-E_{cj}\right)\right),\nonumber \end{eqnarray}
 where the sums and products $j=1,\dots,n$ run over all fragments
of the breakup mode, while the sum $l=1,\cdots,k$ runs over the distinct
fragments and takes into account their multiplicities. Here, $\varepsilon_{0}$
is the excitation energy of the decaying nucleus, $B_{0}$ its binding
energy and $E_{c0}$ is a term associated with the Wigner-Seitz correction
to the Coulomb energy of the system. $V_{n}$ is the volume in which
the momentum states are normalized and is usually defined as \cite{geant}\begin{equation}
V_{n}=\left(1+\chi\right)V_{0}\,,\end{equation}
 where $V_{0}$ is the ground state volume of the decaying nucleus
and the expansion factor $\chi$ is usually taken to be $\chi=1$.
For the fragments, $B_{j}$ is the binding energy of fragment $j$
and $g_{j}$ is its spin multiplicity, while the $E_{cj}$ represent
the remaining Wigner-Seitz corrections to the Coulomb energy, taken
to be \begin{equation}
E_{cj}=\frac{C_{Coul}}{\left(1+\chi\right)^{1/3}}\frac{Z_{j}^{2}}{A_{j}^{1/3}}\,.\end{equation}
 Conservation of nucleon number and charge requires that\begin{equation}
A_{0}=\sum_{j=1}^{n}A_{j}=\sum_{l=1}^{k}N_{l}\, A_{l}\qquad\mbox{and}\qquad Z_{0}=\sum_{j=1}^{n}Z_{j}=\sum_{l=1}^{k}N_{l}\, Z_{l}\,,\end{equation}
 where $Z_{j}$ and $A_{j}$ are the charge and mass number, respectively,
of fragment $j$. The FBM assumes that the fragments are emitted in
their ground states or in (almost) particle-stable excited states.

As the total excitation energy is increased, other particle-unstable
excited states that are long-lived in comparison to the initial decaying
nucleus could also be included and can make significant contributions
to the phase space integral\cite{Rezende}. These can be incorporated
compactly using the densities of excited states of the fragments.
Such an extension of the Fermi breakup integral takes the form \begin{eqnarray}
\omega_{n} & = & \prod_{l=1}^{k}\frac{1}{N_{l}!}\,\left(\frac{V_{n}}{\left(2\pi\hbar\right)^{3}}\right)^{n-1}\int\prod_{j=1}^{n}d^{3}p_{j}\,\delta\left(\sum_{j=1}^{n}\vec{p}_{j}\right)\,\label{genFBM}\\
 &  & \qquad\times\int\prod_{j=1}^{n}\left(\omega_{j}\left(\varepsilon_{j}\right)d\varepsilon_{j}\right)\,\delta\left(\varepsilon_{0}-B_{0}-E_{c0}-\sum_{j=1}^{n}\left(\frac{p_{j}^{2}}{2m_{j}}+\varepsilon_{j}-B_{j}-E_{cj}\right)\right),\nonumber \end{eqnarray}
 where $\varepsilon_{j}$ is the excitation energy of fragment $j$,
$\omega_{j}\left(\varepsilon_{j}\right)$ its density of states and
$B_{j}$ is now its ground-state binding energy. Note that this expression
does not contain the fragment spin multiplicities, $g_{j}$, which
are now assumed to be incorporated in the density of states. For a
particle with no excited states, we have $\omega_{j}\left(\varepsilon_{j}\right)=g_{j}\delta\left(\varepsilon_{j}\right)$.

After rewriting the densities of fragment states in terms of the internal
Helmholtz free energies, defined for fragment $j$ by\cite{Sergio1}\begin{equation}
e^{-\beta_{j}f_{j}^{*}\left(\beta_{j}\right)}=\int_{0}^{\infty}d\varepsilon_{j}e^{-\beta_{j}\varepsilon_{j}}\omega_{j}\left(\varepsilon_{j}\right)\,,\end{equation}
 all but one of the integrals in Eq. (\ref{genFBM}) can be performed
analytically, as is shown in the Appendix. We can then write the density
of final states $\omega_{n}$ as \begin{equation}
\omega_{n}=\frac{1}{2\pi i}\,\int_{c-i\infty}^{c+i\infty}d\beta\,\exp\left[-\beta\left(F_{n}\left(\beta\right)-E_{0}\right)\right]\,,\label{gensmm}\end{equation}
 where \begin{equation}
E_{0}=\varepsilon_{0}-B_{0}\,\end{equation}
 and the total Helmholtz free energy $F_{n}\left(\beta\right)$ has
been defined as \begin{equation}
F_{n}\left(\beta\right)=\sum_{l=1}^{k}N_{l}\left(f_{l}^{*}\left(\beta\right)+f_{l}^{trans}\left(\beta\right)-B_{l}-E_{cl}\right)-\left(f_{0}^{trans}\left(\beta\right)-E_{c0}\right),\label{freee}\end{equation}
 with the sum over fragments replaced by a sum over distinct fragments
times their multiplicities. The translational Helmholtz free energies
are given by\begin{equation}
f_{l}^{trans}\left(\beta\right)=-\frac{1}{\beta}\left[\ln\left(V_{n}\left(\frac{m_{N}A_{l}}{2\pi\hbar^{2}\beta}\right)^{3/2}\right)-\frac{\ln\left(N_{l}!\right)}{N_{l}}\right]\,,\end{equation}
 where we write the mas of fragment $l$ as $m_{l}=m_{N}A_{l}$ and
the mass of the decaying nucleus as $m_{0}=m_{N}A_{0}$, with $m_{N}$
the nucleon mass. We emphasize that we have made no approximations
up to this point. The expression given in Eq. (\ref{gensmm}) is exactly
equivalent to that of Eq. (\ref{genFBM}).

To approximate the final integral, we use the method of steepest descent.
We look for a value $\beta_{0}$ for which\begin{equation}
\left.\frac{d\;}{d\beta}\left(\beta F_{n}\left(\beta\right)\right)\right|_{\beta_{0}}-E_{0}=\left.\left(F_{n}\left(\beta\right)+\beta\frac{dF_{n}}{d\beta}\right)\right|_{\beta_{0}}-E_{0}=0\,.\end{equation}
 Using the relations of the Helmholtz free energy to the entropy and
energy,\begin{equation}
s=-\frac{df}{dT}=\beta^{2}\frac{df}{d\beta}\qquad\mbox{and}\qquad e=f+Ts=f+\beta\frac{df}{d\beta},\end{equation}
 respectively, we find the saddle point condition to be equivalent
to the requirement that energy is conserved,\begin{equation}
\sum_{l=1}^{k}N_{l}\left(e_{l}^{*}\left(\beta_{0}\right)+e_{l}^{trans}\left(\beta_{0}\right)-B_{l}-E_{cl}\right)-\left(e_{0}^{trans}\left(\beta_{0}\right)-E_{c0}\right)=\varepsilon_{0}-B_{0}.\label{econs}\end{equation}
 At the saddle point $\beta_{0}$, the argument of the exponential
is then the total entropy, $S_{n}(\beta_{0})$,\begin{eqnarray}
-\beta_{0}\left(F_{n}\left(\beta_{0}\right)-E_{0}\right) & = & \beta_{0}^{2}\sum_{l=1}^{k}N_{l}\left(\frac{df_{l}^{*}}{d\beta}+\frac{df_{l}^{trans}}{d\beta}\right)-\beta_{0}^{2}\frac{df_{0}^{trans}}{d\beta}\label{entropy}\\
 & = & \sum_{l=1}^{k}N_{l}\left(s_{l}^{*}\left(\beta_{0}\right)+s_{l}^{trans}\left(\beta_{0}\right)\right)-s_{0}^{trans}\left(\beta_{0}\right)\equiv S_{n}\left(\beta_{0}\right)\,.\nonumber \end{eqnarray}
 To complete the evaluation, we must calculate the second derivative
to determine the direction of steepest descent. Since 
\begin{equation}
S_n=-dF_n/dT \qquad\mbox{and}\qquad dS_n/dT=C_{V,n}/T,
\end{equation}
where $C_{V,n}$ is the specific heat of the configuration at constant
volume, we have
\begin{equation}
\frac{d^{2}F_{n}}{dT^{2}} = - \frac{C_{V,n}}{T}\qquad\mbox{and}\qquad 
\frac{d^{2}\;}{d\beta^{2}}\left(\beta F_{n}\left(\beta\right)\right)=T^{3}\frac{d^{2}F_{n}}{dT^{2}} = -C_{V,n} T^2\,. 
\end{equation}
Near the saddle point, we then find, with $T_{0}=1/\beta_{0}$,
\begin{eqnarray}
-\beta\left(F_{n}\left(\beta\right)-E_{0}\right) & \approx & -\beta_{0}\left(F_{n}\left(\beta_{0}\right)-E_{0}\right)-\frac{1}{2}\left.\frac{d^{2}\;}{d\beta^{2}}\left(\beta F_{n}\left(\beta\right)\right)\right|_{\beta_{0}}\left(\beta-\beta_{0}\right)^{2}\\
 & \approx & S\left(T_{0}\right)+\frac{1}{2}C_{V,n}T_{0}^{2}\left(\beta-\beta_{0}\right)^{2}\,.\nonumber \end{eqnarray}
 We thus conclude that the direction of steepest descent is purely
imaginary. The integral then yields\begin{equation}
\omega_{n}=\frac{\exp\left(S_{n}\left(T_{0}\right)\right)}{\sqrt{2\pi C_{V,n}T_{0}^{2}}}\,,\label{FBMdens}\end{equation}
 which is valid as long as the specific heat $C_{V,n}$ is positive.
When one uses the usual SMM approximation to the internal energy\cite{Raul1,Raul2,Raul3,Botvina1,Botvina2,Sergio1}
to evaluate this expression, the specific heat can become negative
when the negative surface term of the internal energy of one of the
fragments dominates the positive bulk term. But in that case, the
fragment can no longer be considered to exist.

As $T_{0}\rightarrow0$, we expect the phase space integral to reduce
to the original form of the FBM, Eq.(\ref{FBM}), for which a well-known
closed-form expression exists, \begin{equation}
\omega_{n}\rightarrow\prod_{l=1}^{k}\frac{1}{N_{l}!}\,\frac{1}{m_{0}^{3/2}}\,\prod_{j=1}^{n}g_{j}m_{j}^{3/2}\,\left(\frac{V_{n}}{\left(2\pi\right)^{3/2}\hbar^{3}}\right)^{n-1}\,\frac{E_{kin}^{3\left(n-1\right)/2-1}}{\Gamma\left(3\left(n-1\right)/2\right)}\,,\qquad\mbox{as}\quad T_{0}\rightarrow0\,,\end{equation}
 where\begin{equation}
E_{kin}=\varepsilon_{0}-B_{0}-E_{c0}+\sum_{j=1}^{n}\left(B_{j}+E_{cj}\right)\end{equation}
 is the total kinetic energy of the fragments. The steepest-descent
approximation does indeed reduce to this closed-form expression, with
all fragments in their ground states, as $T_{0}\rightarrow0$, except
for a multiplicative factor $R\left(3\left(n-1\right)/2\right)$ that
substitutes a Stirling approximation for the gamma function in the
denominator,\[
R\left(h\right)=\Gamma\left(h\right)/\exp\left[\left(h-1/2\right)\ln\left(h\right)-h+\ln\left(2\pi\right)/2\right]\,.\]
 This factor is approximately 1.06 for $n=2$, 1.03 for $n=3$ and
decreases to one as $n$ increases.

\section{Comparison of the FBM and the generalized FBM}

To get an idea of the differences to be expected between a calculation
using the usual FBM and its generalized version, we compare calculations
using the two models here. To set a context, we consider the case
of a 62 MeV proton incident on $^{16}$O, for which double differential
emission data exist \cite{Bertrand}. We describe the initial stage
of the reaction using the Monte Carlo exciton cascade model of Blann
and Chadwick \cite{Blann1,Blann2}, which provides a good description
of the double differential data. This model simulates the pre-equilibrium
stage of a nucleon-nucleus collison by following a cascade of particle-particle
and particle-hole interactions together with particle emission, until
all remaining nucleons have energy smaller than their separation energy
from the residual nucleus. When it is applied to the case of a 62
MeV proton incident on $^{16}$O, it furnishes a $^{16}$O primary
compound nucleus population corresponding to 66\% of the reaction
cross section of approximately 400 mb, with the primary populations
of the $^{15}$O and $^{15}$N compound nuclei accounting for another
32\% of the reaction cross section. The distribution in excitation
energy of the compound populations is extremely broad, with that of
$^{16}$O extending from zero to the center-of-mass energy of 58.4
MeV with a peak at about 50 MeV.

Using this as motivation, we compare the results for the
Fermi breakup of $^{16}$O at an excitation energy of 50 MeV including
1) only the ground states of the fragments, 2) the ground states and
particle-bound states of the fragments and 3) the ground states and
all excited states of the fragments found in the RIPL-2 nuclear level
library \cite{RIPL2}. We note that the calculation including ground
states and particle-bound states is not entirely consistent as it
includes particle-unbound ground states, such as those of $^{5}$He
and $^{8}$Be, which make important contributions to the primary fragmentation
yields. We also point out that the calculation using all excited states
cannot be considered complete, as the discrete level sets of the RIPL-2
library are incomplete at the energies available to the heavier fragments
of several two-body decay modes, in particular, of $n+^{15}$O, $p+^{15}$N
and $\alpha+^{12}$C. Nevertheless, the discrete level sets contain
a sufficient number of levels to clarify the principal differences
to be expected between the models.

\begin{figure}[tb]
\includegraphics[height=12.0cm,angle=270]{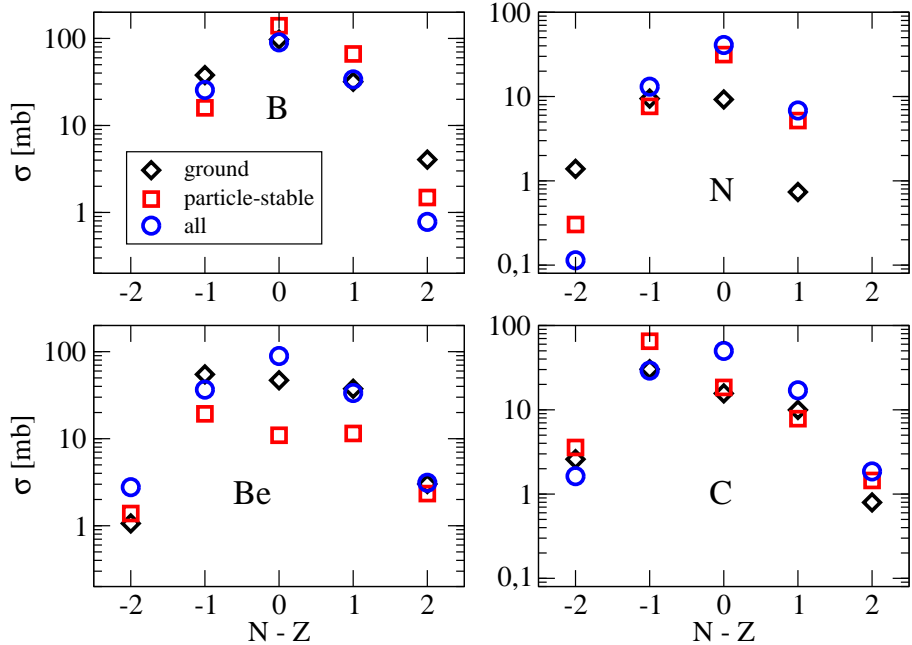}
\caption{ (Color online) Primary fragment production
cross sections of Be, B, C and N isotopes as a function of the neutron
excess, $N-Z$.}
\end{figure}

The calculations were performed using the steepest-descent approximation
in all cases. They furnish 16 two-fragment, 38 three-fragment, 33
four-fragment, 8 five-fragment and 2 six-fragment decay channels.
The contributions of the five- and six-fragment modes are negligible
and the contributions of the four-fragment modes are small. The primary
fragment production cross sections of the isotopes of nitrogen, carbon,
boron and beryllium are shown in Fig. 1 as functions of 
the neutron excess, $N-Z$. The
ground-state-only calculation yields flatter isotopic distributions
than the others, which, with the exception of boron, do not have their
peaks at $N=Z,$ the minimum of the valley of stability. The particle-stable
state calculation tends to be more irregular than the ground-state-only
one. It depresses $^{8}$Be production, due to its lack of particle-bound
states to compete with those of other nuclei. It places the nitrogen
isotope peak at $A=14$, but maintains the carbon isotope maximum
at $A=11,$ as the $n+\alpha+^{11}$C decay mode competes favorably
with the $\alpha+^{12}$C mode due to the limited number of particle-bound
excited states of $^{12}$C. When all excited states are included
in the calculation, all isotopic distributions have their peaks at
the minimum of the valley of stability and roughly reflect the (inverted)
form of the valley. 

The mean primary fragment multiplicity decreases from 2.8 for the
ground-state only calculation to 2.4 for the particle-stable state
one to 2.3 when all excited states are included. Neglecting the four-fragment
or higher modes, which correspond to less than 1\% of the cross section
in all cases, these multiplicities imply that the contribution of
three-fragment modes to the primary fragment distribution is about
80\% in the ground-state only calculation, about 40\% in the particle-stable
calculation and only 25\% in the calculation containing all excited
states. That is, the decay tends from dominance of the three-fragment
modes to dominance of the two-fragment modes as excited states are
taken into account. The decrease in the primary multiplicity is easily
understood, since an increasing portion of the excitation energy remains
in the fragments as excited states are added, rather than being liberated
as a larger number of smaller, lesser-bound fragments. The fact that
the multiplicity decreases from a value close to three to one close
to two is a result of the relatively low excitation energy of the
calculation. At higher excitation energies, the average multiplicity
is larger but follows the same trend - as more excited states are
included, the average primary multiplicity decreases. We note that
the excited fragments of the generalized FBM will, of course, subsequently
decay and increase the net multiplicity. To be consistent, however,
this subsequent decay would also be best described using the generalized
version of the FBM. 

\begin{figure}[tb]
\includegraphics[height=12.0cm,angle=270]{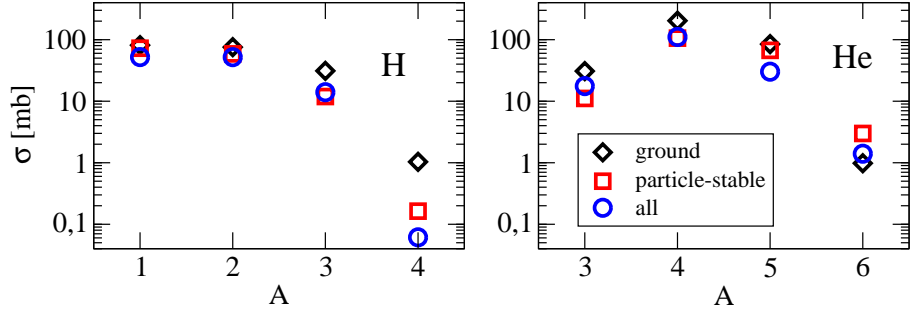}
\caption{ (Color online) Primary fragment production
cross sections of H and He isotopes as a function of mass number.}
\end{figure}

Before concluding this section, we briefly discuss the primary distributions
of hydrogen and helium isotopes, shown in Fig. 2. These remain fairly
stable in relative yield but decrease slightly in magnitude as excited
states are included. The reduction in magnitude is a reflection of
the decrease in multiplicity - the lower multiplicity of the calculations
including excited states is mainly due to decreased emission of H
and He fragments. The similarity in relative yields of these isotopes
is to be expected, as they have either no excited states or very few
excited states at high energy that cannot increase their production
greatly, but are emitted in fragmentations that leave other nuclei
in excited states. An exception to this argument is seen in the production
cross section of $^{4}$H, which comes exclusively from the two-fragment
ground-state decay mode of $^{4}$H+$^{12}$N. As neither of these
residual nuclei have known excited states, this mode is suppressed
as the excited states of other nuclei are included. 

Two further comments should be made in reference to calculations
using the generalized FBM. As alluded to above, sets of known discrete
levels are limited, even for stable nuclei. Continuum level densities
are thus needed to perform reasonably realistic calculations at energies
higher than those we have shown here. Level density parameters that
have been fit to discrete levels and resonance densities can be found
in the RIPL-2 library, but only for $^{20}$F and heavier nuclei.
The parameters for lighter nuclei must be obtained from the extrapolation
of systematics or from theoretical calculations. Second, hot nuclei
have a limiting temperature/excitation energy, above which they no
longer exist. This should be taken into account in FBM calculations,
especially in the case of light nuclei far from stability, where this
temperature/excitation energy is expected to be quite low. One manner
of obtaining estimates to both the level densities and the limiting
temperatures of arbitrary nuclei is through the use of self-consistent
temperature-dependent mean field calculations, such as those we have
recently applied to the SMM \cite{Sergio2}.

\section{From the FBM to the SMM}

The generalized FBM is very similar in form to the SMM used to describe
the fragmentation of highly excited heavy nuclei. We discuss the similarities
and differences of the two models here. To be brief, we refer to the
generalized FBM as simply the FBM in this section. As many versions
of the SMM have been proposed and we do not wish to compare the FBM
to all of them, we will take as our reference the microcanonical SMM
presented in Ref. \cite{Sergio1}, which we will denote simply as
the SMM.

We begin by observing that the SMM uses the Helmholtz free energy
of Eq. (\ref{freee}) to define the entropy of Eq. (\ref{entropy})
and uses the energy condition of Eq. (\ref{econs}) to determine the
configuration temperature $T_{0}$. It then defines the statistical
weight of a configuration as \begin{equation}
w_{n}\Delta\varepsilon_{0}=\exp\left(S_{n}\left(T_{0}\right)\right)\,,\label{SMMdens}\end{equation}
 where $\Delta\varepsilon_{0}$ is a small interval in energy about
the total value $\varepsilon_{0}$. In most cases, the differences
between the statistical weight of the FBM, Eq.(\ref{FBMdens}), and
that of the SMM, Eq. (\ref{SMMdens}) are irrelevant, as the variations
in specific heats and temperatures among configurations are small
compared to the exponential variation with respect to the entropy\cite{Sergio1}.

In fact, although we entered into great detail to calculate the density
of states of an FBM fragment configuration in terms of the quantities
used in the SMM, the association between the two models was already
established once we identified the statistical weight of the FBM as
the density of states. The microcanonical SMM defines the statistical
weight of a fragment configuration in terms of the microcanonical
entropy $S_{n}$, which, in turn, can be defined in terms of the density
of states $\omega_{n}$ as \begin{equation}
S_{n}=\ln\left(\omega_{n}\Delta\varepsilon_{0}\right)\,.\end{equation}
 Thus, the FBM and the SMM use the same physical quantities
in much the same way. They still have their differences, however,
which we discuss next.

A minor distinction between the FBM and the SMM, as they are commonly
used, is their treatment of the normalization volume $V_{n}$. Both
models use a normalization volume larger than the volume of the decaying
nucleus $V_{0}$, with the difference expressed in terms of a multiplicative
factor $\chi$. As mentioned before, calculations with the FBM model
often use a normalization volume twice that of the volume of the decaying
nucleus\cite{geant},\begin{equation}
V_{n,FBM}=\left(1+\chi\right)V_{0}\quad\mbox{with}\quad\chi=1\,.\end{equation}
 SMM calculations normally use a factor of $\chi$ between 2 and 5,
but exclude the volume of the fragments\cite{Raul1,Sergio1},
so that\begin{equation}
V_{n,SMM}=\left(\left(1+\chi\right)V_{0}-\sum_{j=1}^{n}V_{j}\right)=\chi V_{0}\quad\mbox{with}\quad\chi=2\,-\,5.\end{equation}
 As defined here, the volumes in both models are independent of the
fragment configuration and are obviously the same when $\chi=2$ is
used in the SMM. In Ref. \onlinecite{Raul2}, a multiplicity-dependent
volume was used in order to assure that the fragments were formed
outside their respective Coulomb radii. This effectively introduced
a factor $\chi$ that increases with the total excitation energy,
thereby increasing the volume available to the fragments as the excitation
energy increases. Recently, we have used temperature-dependent fragment
volumes $V_{j}$ obtained from self-consistent calculations of the
structure of hot nuclei in the SMM\cite{Sergio2}. In this case, the
normalization volume is temperature and configuration dependent and
no longer reduces to $\chi V_{0}$. We note that the reduction of
the normalization volume due to exclusion of the fragment volumes
can be justified in both the FBM and the SMM, based on considerations
similar to those used to obtain the van der Waals approximation to
the equation of state of a near-ideal gas. Numerical calculations
show the volume reduction to depend on both the masses and the number
of fragments\cite{DasGupta2,Raduta}, but to be reasonably well described
by the approximate form given above.

The most important difference between the two models could be considered
one of philosophy. The FBM is a model of nuclear decay while the SMM
is an equilibrium statistical model whose configurations are identified
with the fragmentation modes of the decaying nucleus. This difference
is reflected in the fact that the SMM considers the configuration
containing only one fragment, the decaying nucleus, that the FBM
does not take into account. This has been justified by characterizing
the SMM decay as \emph{explosive} and contrasting it to the \emph{slower}
compound nucleus (CN) decay, which all residual fragments are assumed
to undergo, including the remaining fraction of the original (one-fragment)
configuration, after the initial fragmentation\cite{Botvina1,Botvina2}.

Unfortunately, neither the FBM nor the SMM furnish decay widths or
lifetimes that could be used to compare their characteristic time
scales with those of CN decay. However, one property accessible in
both the FBM/SMM and the CN decay models is the average energy of
the emitted particles. In the SMM, it has been shown that collective
flow due to radial expansion contributes little to the fragment energies
\cite{Sergio3}. The average relative asymptotic energy of the fragments
of a two-body FBM/SMM decay (assuming fragment volumes independent
of the temperature) can then be taken to be $3T_{0}/2+\tilde{V}_{c}$,
where $\tilde{V}_{c}$ is the energy gained due to the post-emission
Coulomb repulsion of the two fragments. The Weisskopf approximation
to CN emission of a particle of type $c$ (two-body decay) furnishes
a statistical weight that can be written as, \begin{equation}
2\pi\rho_{0}\left(\epsilon_{0}\right)\Gamma_{c}\left(\epsilon_{0}\right)=\int_{0}^{\infty}\, d\epsilon_{c}\, g_{c}\frac{2\mu_{c}\epsilon_{c}}{\pi\hbar^{2}}\sigma_{c}\left(\epsilon_{c}\right)\rho_{c}\left(\varepsilon_{0}-\epsilon_{c}-Q_{c}\right)\label{weisskopf}\end{equation}
 where $\rho_{0}$ and $\rho_{c}$ are the level densities of the
initial and residual nuclei, respectively, $Q_{c}$ is the $Q$-value
of the reaction and $\sigma_{c}(\epsilon_{c})$ is the absorption
cross section for particles of type $c$ incident on the residual
nucleus at energy $\epsilon_{c}$. This implies an average relative
energy for emission of particles of type $c$ of $2T_{c}+V_{c}$,
where $V_{c}$ is the effective Coulomb barrier between the emitted
particle $c$ and the residual nucleus and the temperature $T_{c}$
is very close to the temperature $T_{0}$ obtained from the same two-body
decay in the FBM \cite{Rezende}. Judging from the energy released
by the two-body decay mode, we would thus have to conclude that the
FBM/SMM emission is no more explosive than the CN one. We thus suspect
that the distinction made between FBM/SMM decay modes and those of
the CN is a spurious one. If this is the case, there is no reason
to retain the original one-body configuration in the SMM, to later
decay by CN emission, as this emission is already taken into account
by the FBM/SMM two-body decay modes.

\section{Concluding remarks}

We have shown that the FBM and the microcanonical SMM can be considered
to be essentially one and the same model, if the FBM is generalized
to include excited states and the one-fragment configuration is excluded
from the SMM. The sequential CN decay assumed to occur after fragmentation
in the SMM would then also be described in a more consistent manner
by application of the FBM/SMM itself to the fragments. That is, the
sequential CN decay would be substituted by sequential multifragmentation.
This would naturally resolve the arbitrary division imposed in many
calculations of post-fragmentation decay through sequential two-body
CN decay for heavy nuclei and Fermi breakup for light nuclei. It,
however, has the drawback of replacing well-known and very well-established
expressions for two-body decay, such as the Weisskopf one of Eq. (\ref{weisskopf}),
with the simple FBM/SMM expression. The ideal solution would be an
improved FBM/SMM, in which the two-body decay is described by a Weisskopf-like
expression and $n$-body decay by an appropriate extension of this. We would
then have a consistent model of equilibrium statistical decay, in
which the decay modes are dictated by the available energy and the
characteristics of the system rather than by its modelers. Work in
this direction is in progress.

\begin{acknowledgments}
We would like to acknowledge the CNPq, FAPERJ, FAPESP, the PRONEX
program, under contract No E-26/171.528/2006, and the International
Atomic Energy Agency, under research contract No. 14568, for partial
financial support. This work was supported in part by the National
Science Foundation under Grant Nos.\ PHY-0606007 and INT-0228058.
AWS is supported by the Joint Institute for Nuclear Astrophysics at
MSU under NSF PHY grant 08-22648 and also by NSF grant PHY 08-00026.
\end{acknowledgments}

\appendix

\section{Evaluation of the phase space integral}

We wish to evaluate the final density of states given by the extended
Fermi breakup integral of Eq.(\ref{genFBM}), which is repeated below,
\begin{eqnarray}
\omega_{n} & = & \prod_{l=1}^{k}\frac{1}{N_{l}!}\,\left(\frac{V_{n}}{\left(2\pi\hbar\right)^{3}}\right)^{n-1}\int\prod_{j=1}^{n}d^{3}p_{j}\,\delta\left(\sum_{j=1}^{n}\vec{p}_{j}\right)\,\\
 &  & \qquad\times\int\prod_{j=1}^{n}\left(\omega_{j}\left(\varepsilon_{j}\right)d\varepsilon_{j}\right)\,\delta\left(\varepsilon_{0}-B_{0}-E_{c0}-\sum_{j=1}^{n}\left(\frac{p_{j}^{2}}{2m_{j}}+\varepsilon_{j}-B_{j}-E_{cj}\right)\right).\nonumber \end{eqnarray}
 We begin by using the formal relation between the densities of states
\(\omega_{j}\left(\varepsilon_{j}\right)\) and the corresponding internal
Helmholtz free energies \(f_{j}^{*}\left(\beta_{j}\right)\)\cite{Sergio1}\begin{equation}
e^{-\beta_{j}f_{j}^{*}\left(\beta_{j}\right)}=\int_{0}^{\infty}d\varepsilon_{j}e^{-\beta_{j}\varepsilon_{j}}\omega_{j}\left(\varepsilon_{j}\right)\,.\end{equation}
 We invert the Laplace transform to obtain an expression for the density
of states \(\omega_{j}\left(\varepsilon_{j}\right)\) as\begin{equation}
\omega_{j}\left(\varepsilon_{j}\right)=\frac{1}{2\pi i}\int_{c_{j}-i\infty}^{c_{j}+i\infty}d\beta_{j}\, e^{\beta_{j}\varepsilon_{j}}e^{-\beta_{j}f_{j}^{*}\left(\beta_{j}\right)}\,,\end{equation}
 where \(c_{j}\) is a positive number to the right of all singularities
in the complex plane. Since the density of states \(\omega_{j}\left(\varepsilon_{j}\right)\)
normally grows as \(\exp\left(2\sqrt{a\varepsilon_{j}}\right)\), the
constant \(c_{j}\) in the inverse Laplace transform could be taken
to zero without any effect on the result. (The integral defining \(f_{j}^{*}\left(\beta_{j}\right)\)
converges for any \(\beta_{j}\) with \(\Re\left[\beta_{j}\right]>0\).)
However, to facilitate the evaluation of intermediate results, it
is convenient to leave it free for the moment. Substituting for the
densities, we have\begin{eqnarray}
\omega_{n} & = & \prod_{l=1}^{k}\frac{1}{N_{l}!}\,\prod_{j=1}^{n}\left(\frac{V_{n}}{\left(2\pi\hbar\right)^{3}}\right)^{n-1}\,\int\prod_{j=1}^{n}d^{3}p_{j}\,\delta\left(\sum_{j=1}^{n}\vec{p}_{j}\right)\,\left(\frac{1}{2\pi i}\right)^{n}\,\prod_{j=1}^{n}\left(\int_{c_{j}-i\infty}^{c_{j}+i\infty}\, d\beta_{j}\, e^{-\beta_{j}f_{j}^{*}\left(\beta_{j}\right)}\right)\nonumber \\
 &  & \qquad\times\int_{0}^{\infty}\prod_{j=1}^{n}\left(e^{\beta_{j}\varepsilon_{j}}d\varepsilon_{j}\right)\,\delta\left(\varepsilon_{0}-B_{0}-E_{c0}-\sum_{j=1}^{n}\left(\frac{p_{j}^{2}}{2m_{j}}+\varepsilon_{j}-B_{j}-E_{cj}\right)\right)\,.\end{eqnarray}
 We begin by integrating over the excitation energies. The first integral,
over \(\varepsilon_{1}\), for example, furnishes\begin{eqnarray}
\omega_{n} & = & \prod_{l=1}^{k}\frac{1}{N_{l}!}\,\prod_{j=1}^{n}\left(\frac{V_{n}}{\left(2\pi\hbar\right)^{3}}\right)^{n-1}\,\int\prod_{j=1}^{n}d^{3}p_{j}\,\delta\left(\sum_{j=1}^{n}\vec{p}_{j}\right)\,\left(\frac{1}{2\pi i}\right)^{n}\,\prod_{j=1}^{n}\left(\int_{c_{j}-i\infty}^{c_{j}+i\infty}\, d\beta_{j}\, e^{-\beta_{j}f_{j}^{*}\left(\beta_{j}\right)}\right)\nonumber \\
 &  & \qquad\times\exp\left[\beta_{1}\left(\varepsilon_{0}-B_{0}-E_{c0}-\sum_{j=1}^{n}\left(\frac{p_{j}^{2}}{2m_{j}}-B_{j}-E_{cj}\right)\right)\right]\,\prod_{j=2}^{n}\int_{0}^{\infty}e^{\left(\beta_{j}-\beta_{1}\right)\varepsilon_{j}}d\varepsilon_{j}\,.\end{eqnarray}
 The remaining integrals over \(\varepsilon_{2,}\dots,\varepsilon_{n}\)
can now be performed if we take \(\Re\left[\beta_{1}\right]>\Re\left[\beta_{j}\right]\),
\(j=2,\dots,n\). The result is \begin{eqnarray}
\omega_{n} & = & \prod_{l=1}^{k}\frac{1}{N_{l}!}\,\prod_{j=1}^{n}\left(\frac{V_{n}}{\left(2\pi\hbar\right)^{3}}\right)^{n-1}\,\int\prod_{j=1}^{n}d^{3}p_{j}\,\delta\left(\sum_{j=1}^{n}\vec{p}_{j}\right)\,\left(\frac{1}{2\pi i}\right)^{n}\,\prod_{j=1}^{n}\left(\int_{c_{j}-i\infty}^{c_{j}+i\infty}\, d\beta_{j}\, e^{-\beta_{j}f_{j}^{*}\left(\beta_{j}\right)}\right)\nonumber \\
 &  & \qquad\times\exp\left[\beta_{1}\left(\varepsilon_{0}-B_{0}-E_{c0}-\sum_{j=1}^{n}\left(\frac{p_{j}^{2}}{2m_{j}}-B_{j}-E_{cj}\right)\right)\right]\,\prod_{j=2}^{n}\frac{1}{\beta_{1}-\beta_{j}}\,.\end{eqnarray}
 We next perform the integrals over \(\beta_{j}\), \(j=2,\dots,n\).
To do this, we use the fact that the function \(\exp\left[-\beta_{j}f_{j}^{*}\left(\beta_{j}\right)\right]\)
is well behaved (no singularities) for \(\Re\left[\beta_{j}\right]>0\).
In particular, we note that as \(\Re\left[\beta_{j}\right]\rightarrow\infty\),
\(\exp\left[-\beta_{j}f_{j}^{*}\left(\beta_{j}\right)\right]\rightarrow g_{j}\),
where \(g_{j}\) is the ground-state degeneracy of fragment \(j\). We
can thus close the \(\beta_{j}\) contour to the right, to obtain\begin{equation}
-\frac{1}{2\pi i}\,\int_{c_{j}-i\infty}^{c_{j}+i\infty}d\beta_{j}\,\frac{e^{-\beta_{j}f_{j}^{*}\left(\beta_{j}\right)}}{\beta_{j}-\beta_{1}}=e^{-\beta_{1}f_{j}^{*}\left(\beta_{1}\right)}\,,\end{equation}
 since \(\Re\left[\beta_{1}\right]>\Re\left[\beta_{j}\right]\). After
performing the \(\beta_{j}\) integrals, \(j=2,\dots,n\), we have (taking
\(\beta_{1}\rightarrow\beta\))\begin{eqnarray}
\omega_{n} & = & \prod_{l=1}^{k}\frac{1}{N_{l}!}\,\left(\frac{V_{n}}{\left(2\pi\hbar\right)^{3}}\right)^{n-1}\,\int\prod_{j=1}^{n}d^{3}p_{j}\,\delta\left(\sum_{j=1}^{n}\vec{p}_{j}\right)\label{parFBM}\\
 &  & \qquad\times\frac{1}{2\pi i}\,\int_{c-i\infty}^{c+i\infty}d\beta\,\exp\left[\beta\left(\varepsilon_{0}-B_{0}-E_{c0}-\sum_{j=1}^{n}\left(\frac{p_{j}^{2}}{2m_{j}}+f_{j}^{*}\left(\beta\right)-B_{j}-E_{cj}\right)\right)\right]\,.\nonumber \end{eqnarray}
 We next perform the integrals over the momenta. Using the integral
representation of the \(\delta\) function, we have\begin{eqnarray}
\int\prod_{j=1}^{n}d^{3}p_{j}\,\delta\left(\sum_{j=1}^{n}\vec{p}_{j}\right)\exp\left[-\beta\sum_{j=1}^{n}\frac{p_{j}^{2}}{2m_{j}}\right] & = & \frac{1}{\left(2\pi\right)^{3}}\int d^{3}x\,\prod_{j=1}^{n}\int d^{3}p_{j}\,\exp\left[i\vec{x}\cdot\sum_{j=1}^{n}\vec{p}_{j}-\beta\sum_{j=1}^{n}\frac{p_{j}^{2}}{2m_{j}}\right]\nonumber \\
 & = & \prod_{j=1}^{n}\left(\frac{2\pi m_{j}}{\beta}\right)^{3/2}\frac{1}{\left(2\pi\right)^{3}}\int d^{3}x\,\exp\left[-\frac{m_{0}}{\beta}\frac{x^{2}}{2}\right]\nonumber \\
 & = & \prod_{j=1}^{n}\left(\frac{2\pi m_{j}}{\beta}\right)^{3/2}/\left(\frac{2\pi m_{0}}{\beta}\right)^{3/2},\end{eqnarray}
 where \(m_{0}=\sum_{j=1}^{n}m_{j}\). We rewrite the full integral
as\begin{eqnarray}
\omega_{n} & = & \prod_{l=1}^{k}\frac{1}{N_{l}!}\,\frac{1}{2\pi i}\,\int_{c-i\infty}^{c+i\infty}d\beta\:\prod_{j=1}^{n}\left(V_{n}\left(\frac{m_{j}}{2\pi\hbar^{2}\beta}\right)^{3/2}\right)/\left(V_{n}\left(\frac{m_{0}}{2\pi\hbar^{2}\beta}\right)^{3/2}\right)\\
 &  & \qquad\qquad\qquad\times\,\exp\left[\beta\left(\varepsilon_{0}-B_{0}-E_{c0}-\sum_{j=1}^{n}\left(f_{j}^{*}\left(\beta\right)-B_{j}-E_{cj}\right)\right)\right]\,.\nonumber \end{eqnarray}
 Combining the factors resulting from the momentum integrals with
those due to the fragment multiplicities, we define the translational
Helmholtz free energies as\begin{equation}
f_{l}^{trans}\left(\beta\right)=-\frac{1}{\beta}\left[\ln\left(V_{n}\left(\frac{m_{N}A_{l}}{2\pi\hbar^{2}\beta}\right)^{3/2}\right)-\frac{\ln\left(N_{l}!\right)}{N_{l}}\right]\,,\end{equation}
 where we recognize that \(m_{l}=m_{N}A_{l}\) and \(m_{0}=m_{N}A_{0}\),
with \(m_{N}\) the nucleon mass. We can then define the total Helmholtz
free energy as \begin{equation}
F_{n}\left(\beta\right)=\sum_{l=1}^{k}N_{l}\left(f_{l}^{*}\left(\beta\right)+f_{l}^{trans}\left(\beta\right)-B_{l}-E_{cl}\right)-\left(f_{0}^{trans}\left(\beta\right)-E_{c0}\right),\end{equation}
 where we have replaced the sum over the fragments with a sum over
distinct fragments times their multiplicities, and write the density
of final states as \begin{equation}
\omega_{n}=\frac{1}{2\pi i}\,\int_{c-i\infty}^{c+i\infty}d\beta\,\exp\left[-\beta\left(F_{n}\left(\beta\right)-E_{0}\right)\right]\,,\end{equation}
 where we have defined \begin{equation}
E_{0}=\varepsilon_{0}-B_{0}\,.\end{equation}
 Up to this point, we have made several definitions but no approximations.
This expression is exactly equivalent to that of Eq. (\ref{genFBM}).

\end{document}